\documentclass[
 final,            
sort&compress %
  ]{aipproc}
\usepackage{natbib}
\setcitestyle{numbers}

\layoutstyle{8x11double}
\usepackage{indentfirst,multirow}
\newcommand{\UW}[0]{University of Washington}
\newcommand{\ped}[0]{potential energy diagram}
\newcommand{\peds}[0]{potential energy diagrams}
\newcommand{\Ped}[0]{Potential energy diagram}

\newcommand{\ourAddress}[0]{Department of Physics, University of Washington, Seattle, WA 98195-1560}
\newcommand{\PEG}[0]{Physics Education Group}

\newcommand{\studentStatement}[1]{\emph{``#1''}}
\newcommand{\standardFigureWidth}[0]{0.9\columnwidth}
\newcommand{\figref}[1]{Fig.~\ref{fig:#1}}
\newcommand{\difficultyHeading}[1]{\noindent\textbf{#1}}

\newcommand{\vs}[0]{\emph{vs.}}

\newcommand{\vol}[1]{\textbf{#1}}

\begin{document}

\title{Examining student ability to interpret and use potential energy diagrams for classical systems}

\classification{01.40.Fk, 45.20.dh}
\keywords      {physics education research, student understanding, potential energy, potential energy diagram}

\author{Brian M. Stephanik}{
  address={\ourAddress}
}

\author{Peter S. Shaffer}{
  address={\ourAddress}
}


\begin{abstract}
	The Physics Education Group at the \UW\ is examining the extent to which students are able to use graphs of potential energy \vs\ position to infer kinematic and dynamic quantities for a system.
	The findings indicate that many students have  difficulty in relating the graphs to real-world systems. 
	Some problems seem to be graphical in nature (\emph{e.g.,} interpreting graphs of potential energy \vs\ position as graphs of position \vs\ time). 
	Others involve relating the graphs to total, kinetic, and potential energies, especially when the potential energy is negative. 
	The results have implications beyond the introductory level since graphs of potential energy are used in advanced courses on classical and quantum mechanics.
\end{abstract}

\maketitle

\date{Mon. July 4th, 2011}


\section{Introduction}

  Potential energy offers a powerful framework with which to characterize and understand complex systems.
  However, this concept also introduces a significant conceptual hurdle for many students. 
  Some specific difficulties with this concept have been documented in prior studies~\cite{Lindsey2009,Lindsey2011,Heuvelen2001,Singh2003}.
  Other papers describe instructional strategies for teaching energy and potential energy~\cite{Sherwood1983,Arons1999,JewettSystems2008}.

  This paper describes preliminary results from an investigation by the \PEG\ at the \UW\ (UW) into student thinking about \peds\ (\emph{i.e.}, graphs of potential energy of a system \vs\ position of a particle). 
  The focus is on the ability of introductory students to use \peds\ to determine kinematic quantities and to reason about total, kinetic, and potential energies.
  
  This investigation was motivated, in part, by research that our group and others have been conducting into student understanding of basic quantum mechanics~\cite{ExtensiveReferencesQM}, a topic that is increasingly being covered in introductory physics courses.
  We are interested in probing the extent to which students understand the underlying classical analogues and in using the results to guide the design of instructional materials~\cite{tutorials}. 
  A goal is to help students be able to compare and contrast the predictions of the two models. 
  As part of this investigation, we had designed questions that asked students to draw potential energy diagrams as a step in helping them sketch the corresponding classical probability distributions. 
  We found, however, that many students had difficulty with this first step, even for simple systems~\cite{Bao}. 
  This observation led us to examine the ability of introductory students to use potential energy diagrams in classical mechanics.

\section{Context for investigation}

	This investigation has involved more than 500 students at the UW in the three-quarter introductory calculus-based sequence for scientists and engineers: Phys~121 (mechanics), 122 (E\&M) and 123 (waves, optics, modern physics, and quantum mechanics). 
	Most of the students were enrolled in the regular sections, but some were in an honors version that covers a greater amount of material in greater depth. (Table~\ref{tab:context} summarizes the student populations.)
	The textbooks for these courses include \peds~\cite{Tipler,Moore}, although the time spent in lecture on this topic varied by instructor.
	In each course, students had completed the relevant lecture and tutorial~\cite{tutorials} instruction on mechanical energy by the time the questions were administered.

  \section{Questions used for research}
  
\begin{figure}[tbp]
    		\resizebox
			{\standardFigureWidth}
			{!}
			{\includegraphics{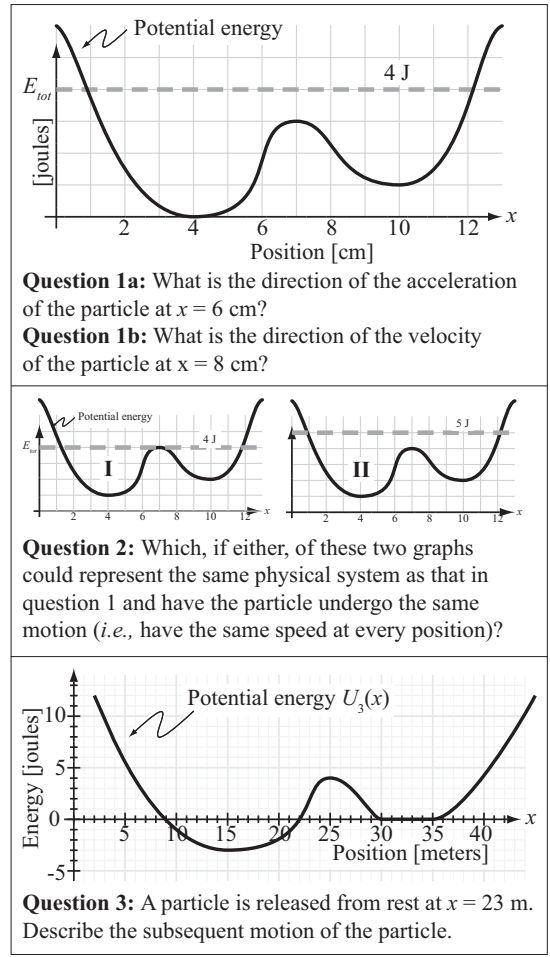}}
		\caption{Examples of questions administered to students.}
		\label{fig:questions}
    \end{figure}

	Several different questions were administered to students in a variety of formats, including individual student interviews and multiple choice exams. 
	(Table~\ref{tab:context} summarizes the various formats.)
	Most questions were designed to probe student ability to relate \peds\ of classical systems to real-world motions.
	In each case, students are asked to consider a particle that is part of a one-dimensional system in which the energy can be treated as consisting only of potential energy of the system and translational kinetic energy of the particle.
	A few examples (questions 1--3) are shown in \figref{questions}.
	
	In questions~1a and~1b, students are shown a \ped\ and asked about the directions of the acceleration and velocity of the particle when it is at a particular position. 
	To answer, they can recognize that the kinetic energy is given by the difference between the total and potential energies.  
	Thus, if the particle were traveling in the positive $x$--direction it would be slowing down, while if the particle were traveling in the negative $x$--direction it would be speeding up. 
	Both motions correspond to acceleration in the negative $x$--direction. 
	Alternatively, students can use the relation $F=-dU/dx$ and Newton's second law. 
	The direction of the velocity of the particle cannot be determined.

	In question~2, students are shown two additional graphs and asked which, if either, can represent the same motion as the original graph. 
	To answer, students can recognize that only graph~II has the same kinetic energy at each point as the original graph. 
	Thus, only graph~II can correspond to the same motion.
	
	\begin{table}
  	\begin{tabular}{llc}
		\hline
		\tablehead{1}{l}{b}{Population}
		 &\tablehead{1}{l}{b}{Question format}
		 &$N$\\
		\hline
		121 & Multiple choice final exam & $165; 183$\\ 
		123 & Individual student interviews & $8$\\
		123 & Online, ungraded quiz & $152$\\
		123 Honors & Written, ungraded quiz & $34$\\
		\hline
	\end{tabular}
	\caption{Student populations and the formats  of the questions that were administered.}
	\label{tab:context}
  \end{table}
	
	In question~3, which was used only during the interviews, students are given a \ped\  and told the particle is released from rest at $x=23$~m. 
	They are then asked to describe the subsequent motion of the particle. 
	In order to answer, students can recognize that since the particle is released from rest, the total energy is equal to the potential energy at that position (\emph{i.e.,} 2~J). 
	Since the slope is positive at the starting position there is a force in the negative direction. Thus, the particle will start to move in that direction.
	Students can then use conservation of energy to determine that at $x=15$~m the particle would obtain a maximum kinetic energy of 5~J $(=2~\textrm{J}-(-3~\textrm{J}))$ before turning around at $x=7$~m, where the kinetic energy is again equal to zero.
    
  \section{Identification of specific difficulties}

   Many students had difficulty in answering each of the questions discussed in the previous section. 
   Some of the problems seemed to be associated with their ability to extract kinematic information from a \ped.  
   Others were related to student ideas about total energy or about potential energy, especially when it is negative. 
   In this section, we discuss some of the more common difficulties. 
   The level of instruction and the formats of the questions varied significantly, therefore we do not compare the percentages of correct and incorrect responses among the various populations.
    
  \subsection{Difficulties related to kinematic quantities}
  
  Many of the errors that students made on each of the questions seemed to be related to their ability to infer kinematic quantities from \peds.  
  These are discussed below.  
  Difficulties related to student understanding of kinematic concepts (\emph{e.g.,} confusing velocity and acceleration) are not discussed since they are documented extensively elsewhere~\cite{Trowbridge1981}.
  
  \difficultyHeading{Belief that \peds\ represent motion only in the positive \emph{x}--direction}
  
	Early in the investigation, we asked question~1a, about the direction of acceleration of the particle, but not the corresponding question about velocity (question~1b).
	Many of the explanations were based explicitly on motion only in the positive $x$--direction. 
	For example, \studentStatement{[the acceleration is in the] negative-x direction. It is moving in $+x$ direction, but gaining Potential E, [therefore] losing KE, [e.g.,] slowing down.} 
	Although students often obtained the correct answer, many seemed to believe that \peds\ only represent motion in a single direction.

	In order to probe student thinking in greater detail, we later added question~1b. 
	On a version given in the waves and optics course, about 15\% of the students stated explicitly that the particle travels only in the positive direction. 
	The reasoning was often circular (\emph{e.g.,} the particle speeds up since the kinetic energy increases, therefore the particle must be moving in the positive $x$--direction).
	This line of reasoning is consistent with the results on a multiple-choice version given in the mechanics course in which about 55\% of the students gave a similar answer.

  \difficultyHeading{Tendency to treat potential energy \vs\ position as position \vs\ time}
  
   Between 10\% and 30\% of the students found the velocity of the particle from the slope of the \ped\ or the acceleration from the second derivative.
   Typical explanations included \studentStatement{[the acceleration is zero since the] double derivative of the graph at $x=6$ is zero.} 
   Some students also described the local minima of \peds\ as turn-around points, consistent with interpreting the slope as the velocity.
   
   In most cases, these responses did not seem to be due simply to students misreading the question or the labels on the axes.
   On the written questions and in the interviews, students sometimes switched between correct and incorrect interpretations.
   
\begin{figure}
    		\resizebox
			{0.5\columnwidth}
			{!}
			{\includegraphics{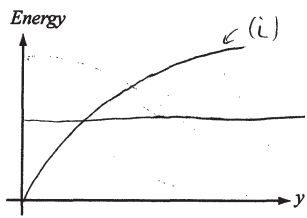}}
		\caption{\Ped\ drawn by a student for an Earth-ball system near the surface of Earth.}
		\label{fig:mghCurved}
    \end{figure}

   It is interesting to note that a similar difficulty arose in a sophomore-level quantum mechanics class (not included in Table~\ref{tab:context}). 
   These students were asked a different question in which they had to sketch a \ped\ for an Earth-ball system near the surface of Earth. 
   Roughly 20\% of the students drew curved graphs as shown in \figref{mghCurved}. 
   (The flat line represents total energy.)
   A common justification was \studentStatement{[t]he ball starts with a potential energy of mgh. 
   As it falls, it accelerates due to gravity and the value h of mgh decreases at an increasing rate.}
   These students appeared to be conveying information about the increasing rate of change of potential energy with respect to time through the slope of the graph.

   \subsection{Difficulties related to\\ negative potential energy}
   Some of the questions used in this study involve \peds\ in which the potential energy is negative in some regions.
   Roughly 35\% of the students struggled in making sense of the negative values.
   Many of the errors seemed to reflect a lack of understanding of the arbitrary choice of reference value for potential energy~\cite{referenceValue} as well as the general relationship between the total, kinetic, and potential energies.
   Two common difficulties are discussed.

   \difficultyHeading{Belief that potential energy cannot be negative}
   
     Many students seemed to believe that potential energy cannot be negative.
     This idea was elicited, for example, in question~3, which was used during the interviews. 
     Some students simply stated that the \ped\ was not possible: \studentStatement{... you cannot have negative potential energy in a system.} 
     Others argued that the expressions $mgh$ and $\frac{1}{2}kx^2$ are always positive. 
     (For these students, $h$ was strictly a positive quantity.)
     They did not seem to recognize that these expressions are a result of a particular reference value for potential energy.
     
     Some students who had studied quantum mechanics used a different argument. 
     About 10\% of the students on a sophomore-level written final exam argued that \studentStatement{classically, there is no way to achieve a negative potential energy.}
     These students regarded negative potential energy as a feature unique to quantum mechanics~\cite{Bao}.
    
    \difficultyHeading{Belief that kinetic energy cannot exceed total energy}
    
    Some students seemed to recognize that potential energy can be negative but had difficulty in relating the total, kinetic, and potential energies when the potential energy was negative.
    For example, one student on question~3 stated:
    \studentStatement{[Y]ou placed it there and let it go, which tells me you didn't give it any kinetic energy at first. So I actually know all of the energy [...,] it's 2~[J]. And at 23~[m] it's 2~[J] of potential energy and 0~[J] of kinetic energy. So it's only going to move until down at 15~[m] and it has 2~[J] of kinetic, and back up to 2~[J] of potential.}
    This student incorrectly stated that the maximum kinetic energy of the particle is 2~J rather than 5~J. 
    This response and many others revealed a strong belief that the numeric value for the kinetic energy cannot exceed the numeric value for the total energy. 
    Common justifications included \studentStatement{energy cannot be created from nothing} and \studentStatement{the total energy of the system encompasses its kinetic energy as well as its PE.} 
    These students did not seem to understand how to interpret the relationship between the numeric values for the total, kinetic, and potential energies when the potential energy was negative.

   \subsection{Difficulties related to total energy}
   
   Some of the student responses revealed insight into their thinking about the total energy of a system. 
   This was particularly the case for question~2, which asked students to identify two different \peds\ that might correspond to the same motion. 
   Between 10\% and 25\% of student answers were consistent with the two ways of reasoning described below.

   \difficultyHeading{Tendency to associate the motion of the particle in a given system with only the total energy}
   
  In comparing the potential energy diagrams in question~2, some students focused only on the total energy. 
  For example, one student, who chose graph~I, stated, \studentStatement{[t]he total energy would [need] to be the same as the previous graph or else the motion of the particle is not the same ...}
  This response and others suggested a tendency to associate the motion of a particle with only the total energy of the system. 
  These students failed to recognize that equal shifts in the total and potential energies could represent the same motion.
   
    \difficultyHeading{Misapplication of conservation of energy}
    
   In answering question~2, many students based their explanations on conservation of energy. 
   For example, \studentStatement{[the system] would have to have the same total energy because energy cannot be created or destroyed ...} or \studentStatement{by the law of con[s]ervation of energy, all energy is conserved.}
   These students did not seem to realize that the arbitrary choice of reference value for potential energy results in an arbitrary total energy.
   They instead used conservation of energy, which states that the total energy does not change with time, to account for differences due to different reference values for potential energy.

  \section{Discussion}
  
    This preliminary investigation has revealed a variety of difficulties that students encounter in interpreting potential energy diagrams.  
    These include determining kinematic quantities as well as relating the total, kinetic, and potential energies, especially when the potential energy is negative. 
    Underlying many of the responses is a failure to understand how different reference values for potential energy do and do not impact the formal description of the energy and motion of a particle.
    Moreover, instruction on more advanced topics may result in additional complications, such as a belief that negative potential energy is only allowed for quantum mechanical systems. 
    
	The results of this research have implications for instruction.  
	For example, on problems involving gravity near the surface of Earth, potential energy is commonly chosen to be zero at the lowest point of the motion of a particle. 
	This choice can result in convenient statements such as ``all of the energy is kinetic at the bottom,'' but may hide difficulties that can arise when the potential energy is negative, which is common or required of many systems.
	The prevalence of the errors and their persistence to the sophomore level, together with observations of students during the interviews, suggest that the underlying difficulties are strongly held and are not likely to be easily addressed.
    There is a need for additional research to probe student thinking in greater detail and to identify instructional strategies that are effective at helping them deepen their understanding of the abstract concept of energy.


\begin{theacknowledgments}
  The authors would like to thank current and past members of the Physics Education Group and the faculty members who welcomed this research into their classes. This research would not have been possible without the support of the National Science Foundation under grants DUE-0618185 and DUE-1022449.
\end{theacknowledgments}


\bibliographystyle{aipproc}   


\begin{thebibliography}{99}
\bibitem{referenceValue} We use the phrase ``reference value for potential energy'' to refer to the choice of value for potential energy at an arbitrary position, which then determines the potential energy values for all other positions.
\end{thebibliography}


\end{document}